\documentclass{article}
\usepackage{commath}
\usepackage{mathptmx,graphicx}
\usepackage[utf8x]{inputenc}
\usepackage{bm}
\usepackage{braket}
\usepackage[T1]{fontenc} 
\usepackage{float}
\usepackage{amsmath,amsfonts,amssymb,eqnarray}  
\usepackage{color, soul}
\usepackage{graphicx}  
\usepackage{dblfloatfix} 
\usepackage{array,booktabs}
\usepackage[pdftex,hypertexnames=false]{hyperref} 
\usepackage[affil-it]{authblk}
\usepackage[font={small}]{caption}
\usepackage{cite}
\usepackage{braket}

\title{Indistinguishable photons on demand from an organic dye molecule}

\author{Pietro Lombardi$^{1\dag}$}
\author{Maja Colautti$^{1}$}
\author{Rocco Duquennoy$^{2}$}
\author{Ghulam Murtaza$^{1}$}
\author{Prosenjit Majumder$^{1}$}
\author{Costanza Toninelli$^{1,3}$*}

\affil{$^1$ \small National Institute of Optics (CNR-INO), Largo Enrico Fermi 6, Firenze 50125, Italy}
\affil{$^2$ \small Dipartimento di Fisica e Astronomia, Universit{\`a} degli Studi di Firenze, Via G. Sansone 1, Sesto Fiorentino 50019, Italy}
\affil{$^3$\small European Laboratory for Non-Linear Spectroscopy (LENS), Via Nello Carrara 1, Sesto F.no 50019, Italy}

\affil{$^\dag$ \small lombardi@lens.unifi.it}
\affil{$^*$ \small toninelli@lens.unifi.it}
\date{\today}

\begin{document}

\maketitle

\begin{abstract}
Single molecules in solid-state matrices have been proposed as sources of single-photon Fock states back 20 years ago. Their success in quantum optics and in many other research fields stems from the  simple recipes used in the preparation of samples, with hundreds of nominally identical and isolated molecules. Main challenges as of today for their application in photonic quantum technologies are the optimization of light extraction and the on-demand emission of indistinguishable photons. We here present Hong-Ou-Mandel experiments with photons emitted by a single molecule of dibenzoterrylene in an anthracene nanocrystal at 3 K, under continuous wave and also pulsed excitation. A detailed theoretical model is applied, which relies on independent measurements for most experimental parameters, hence allowing for an analysis of the different contributions to the two-photon interference visibility, from residual dephasing to spectral filtering. 
\end{abstract}

\section{Introduction}
Two-particle interference is one of the most striking consequences of quantum physics. After the first seminal experiments with photons\cite{abram1986direct,hong87,shih1988new,rarity1989fourth}, the effect has been demonstrated also for massive particles and quasi-particles such as electrons\cite{liu1998quantum,neder2007interference}, plasmons\cite{heeres2013quantum,fakonas2014two} and atoms\cite{kaufman2014two,lopes2015atomic}. Roy Glauber well captured this phenomenon intuitively within a generalized picture of interference, affecting all measurement outcomes which result from multiple indistinguishable stories\cite{Glauber2006}. Two-particle interference though does not have any classical counterpart, not even in the case of light\cite{paul1986interference}. Indeed two-photon interference (TPI) is at the heart of many quantum technologies (see Ref. \cite{Bouchard2021}), potentially outperforming classical computation schemes\cite{knill01,obrien07}, offering physics-protected information\cite{Sangouard2011}, simulating highly complex physical systems\cite{Sparrow2018,abajo02,Angelakis2017} and enhancing the sensitivity of precise measurements\cite{dowling2008quantum,giovannetti11}. The so-called Hong-Ou Mandel (HOM) interferometer, originally developed to measure small time shifts\cite{hong87,shih1988new} nowadays reaching attosecond resolution \cite{Lyons2018}, is a well known experimental scheme unveiling TPI. Essentially it consists of two single-photon detectors at the output ports of a beam splitter (BS). Due to the bosonic nature of photons, in case of two indistinguishable photons impinging separately on the BS ports, interference forces them to coalesce at one of the outputs, suppressing the rate of simultaneous coincidences and hence yielding the two-photon path-entangled state\footnote{The N=2 NOON state with $+$ sign is obtained in the specific case of a beam splitter which is symmetric in the phase accumulated under reflection from both sides.} $\frac{1}{\sqrt{2}}(\ket{2,0}+\ket{0,2})$. In order to enable effective quantum interference, the involved particles have to be prepared in the very same quantum state, so as to determine indistinguishability among different possible events with the same output state. It is also well established that quantum emitters in the solid state play a major role in the generation of single photons on a deterministic basis\cite{Aharonovich2016}, which is key to the preparation of more complex states of light that are useful for many protocols in quantum technologies. However, the ability of such systems to provide photons in the same quantum state remains one of the most elusive requirements to meet, in particular in the sub-micrometric environment of integrated photonics\cite{Liu2018,Jantzen2016}. 

Single molecules of polyaromathic hydrocarbons (PAH) in suitable host matrices are known for emitting with high quantum efficiency in very narrow (few-tens-of-MHz wide) and stable zero phonon lines (ZPL). Molecular quantum emitters can also be integrated inside complex and hybrid photonic structures, yielding almost $100\%$ collection efficiency \cite{Lee2011}, $99\%$ light extinction \cite{Wang2019} and allowing for on-chip photon manipulation \cite{Lombardi2017,Rattenbacher2019}, (for a complete review see e.g. Ref.\cite{Toninelli2020}).
Very promising results have been reported in HOM interference experiments with single-photon streams, obtained from molecular emitters after continuous wave (CW) laser pumping \cite{kiraz05,Trebbia2010indistinguishable,Rezai2018,Rezai2019}, including the case of photons emitted by remote molecules \cite{Lettow2010a}. However, triggered single-photon pulses appears necessary in any kind of logic operation with photons. In this work we demonstrate that highly indistinguishable single photons can be generated on demand by single PAH molecules under pulsed operation. Notably, the results are obtained for non-resonant pumping of the ZPL and without coupling to optical cavities. 
Hong-Ou-Mandel experiments are reported, based on triggered single photons from single dibenzoterrylene (DBT) molecules in anthracene nanocrystals\cite{pazzagli2018,lombardi2020molecule}.
\begin{figure*}
\includegraphics[width=\textwidth]{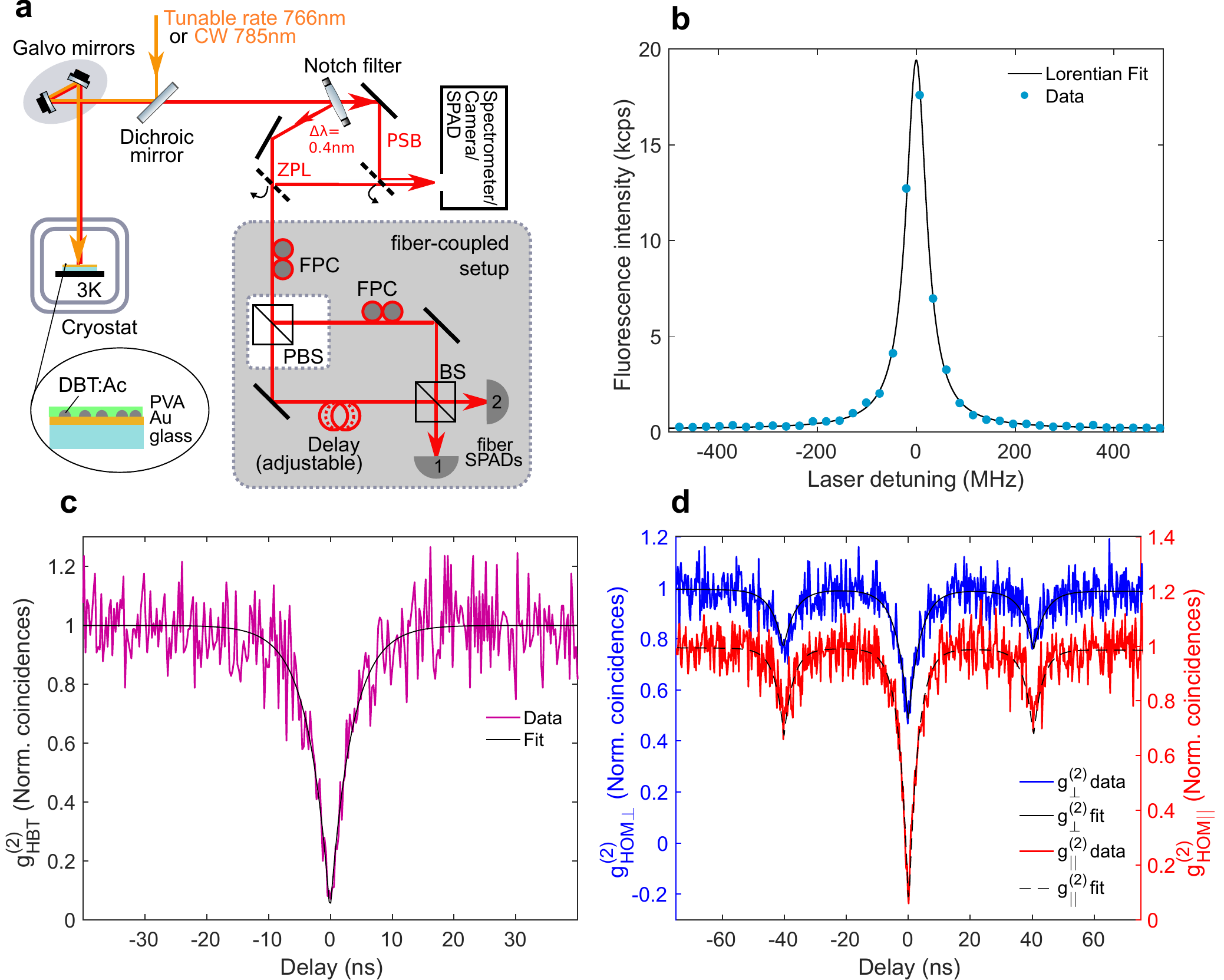}
\caption {\textbf{a) Sketch of the experimental setup} By means of a notch filter, the molecule emission collected in an epifluorescence-microscope configuration is separated in two parts containing the ZPL and the PSB. Both components can then be analyzed with a spectrometer and a free-space-coupled single photon avalanche diode (SPAD). The ZPL component can also enter the fibered circuit for probing purity and indistinguishability of the photon wave packet. \textbf{b) Excitation spectroscopy} Fluorescence intensity in the PSB as the frequency of a resonant laser is scanned accross the ZPL. A lorentian fit to the data yields a FWHM of $55.1\pm0.5$ $MHz$. \textbf{c) Purity of the photon stream under CW operation} Histogram of the relative arrival time between the two SPADs in HBT configuration (no photons in the delay line arm). Solid line is a single exponential fit to the data, yielding $g^{(2)}(0)=0.03\pm0.02$ and $\tau_1=3.5\pm0.2$ $ns$. \textbf{d) HOM results in CW operation} Histogram of the relative photon arrival times in the HOM configuration. Results for parallel (red) and orthogonal (blue) polarizations are superimposed with shifted y-axis scale for clarity. Solid and dashed lines are fits to the data with expression \eqref{HOMcw}.}
\label{fig1}
\end{figure*}

In order to model the outcome of the HOM measurements, we independently characterize also the single-photon purity, the emission linewidth, the molecule lifetime and the setup parameters. The model that is briefly outlined in the Supplementary Information (SI), correctly describes our data and suggests that the main limitation to the observed raw-data visibility is to be found in the residual dephasing observed at 3K, which is currently our lowest temperature of operation. Upon cooling down to 1.4 K an intrinsic wave-function overlap of up to $95\%$ can be estimated. 

 In Fig.\ref{fig1}a the experimental setup used for characterization and for the HOM measurements is outlined. Single-molecule fluorescence is excited and collected in an epi-fluorescence microscope through a 0.7-N.A. objective, using either a tunable CW laser ($\lambda=783.5$ $nm$) or a pulsed laser with central wavelength $\lambda=766.0$
$nm$, adjustable repetition rate and $50$-ps long pulses (Picoquant LDH-D-FA-765L), that can be operated in CW as well. A LongPass filter (Semrock LP02-785RE) is employed for rejection of the back-scattered pump light. The ZPL component at $783.5$ $nm$ is separated from the phonon sideband (PSB) exploiting a $0.4$ $nm$-wide reflective Notch filter (OptiGrate BNF-785-OD4-12.5M), and coupled into a single-mode fiber. In order to mimic two independent sources, the photon stream is
splitted in two arms by a polarizing beam splitter cube (PBS), with a fiber polarization control in front (FPC, Thorlabs FPC023). One of the paths is then delayed by $\Delta t \simeq 40$ $ns$ ($8$ $m$-long fiber), much longer than the excited state lifetime ($\simeq 4$ $ns$), as required to avoid temporal correlations between the photons in the two arms. 
Quantum interference in the train of single photon pulses is then
measured by an unbalanced fiber-based Mach-Zender interferometer (MZI). In particular, a FPC unit is added on one path to switch between parallel and orthogonal configuration, before the 
MZI 
is closed on a fibered $50/50$ BS (Thorlabs TN785R5A2). The two output ports are finally connected to a couple of single-photon counting modules (Excelitas SPCM-NIR-14).
When one of the interferometer arms is blocked, the setup operates in the so-called Hanbury-Brown \& Twiss\cite{brown1956correlation} (HBT) configuration, allowing for the characterization of the photon statistics.
Instead, when the photon flux is equally distributed on the two arms, we can probe the photon indistinguishability through HOM interference.
In Fig.\ref{fig1}c and d, the histogram of the inter-photon arrival times for small time delays, normalized to the background average value, is reported for photon streams collected in the HBT and HOM configuration, respectively (saturation parameter $s\simeq 0.2$). In both cases, the dataset well reproduces the second order correlation function $g^2(\tau)$, probing purity and indistinguishability of the photon stream. From the best fit to the data in Fig.\ref{fig1}c with equation $g^{(2)}_{HBT}(\tau)=(1-b*\exp{(-|\tau|/\tau_{HBT})})$, we obtain $g^{(2)}_{HBT}(0)=0.03 \pm 0.02$.

Moreover, the effect of non-classical interference is clearly visible in panel d around zero time delay, where the data set for parallel polarization is well below the one for photons which are made distinguishable in polarization, i.e. to say orthogonal. The other two dips occurring $\Delta t$-away from the main one correspond to the suppressed probability of having two photons closer than the emitter lifetime, in the case of one photon being transmitted/reflected when the other is reflected/transmitted. From the best fit to this measurement with the following equation, derived from Ref.\cite{patel2008postselective} in the low saturation regime

\begin{eqnarray}
    g_{HOM}^{(2)}(\tau)=&&2\abs{r}^2\abs{t}^2g(\tau)+\left[\abs{t}^4g(\tau-\Delta t) + \abs{r}^4g(\tau+\Delta t)\right]\nonumber\\&&
\times(1-V e^{-\abs{\tau}/\tau_\parallel}) 
\label{HOMcw}
\end{eqnarray}
with $g(t)=g^{(2)}_{HBT}(t)$ and $V=0$ for orthogonal polarizations, we deduce $\Delta t=40.3\pm 0.2$ $ns$ and the unbalance between reflectance $\abs{r}^2$ and transmittance $\abs{t}^2$ of the second beam splitter to be $\abs{r}^2/\abs{t}^2 = 1.10 \pm 0.07$. The latter estimation is in agreement with direct measurements implemented with laser light at $783.5$ $nm$, from which we evaluate both the overall losses of the fiber connector + BS ( $0.06 \pm 0.01$), and $\abs{r}^2$ and $\abs{t}^2$ values to be $0.50 \pm 0.01$  and $0.44 \pm 0.01$, respectively. 
These parameters will be used in analysing the data for pulsed excitation, as they can be considered unchanged. 

The degree of coherence of the molecule emission can be estimated as $\abs{g^{(1)}(0)}^2=\frac{g_{\bot}^{(2)}(0)-g_{\parallel}^{(2)}(0)}{g_{\bot}^{(2)}(0)}=0.89 \pm 0.06$. 
This value results compatible with an estimate of the first order coherence from the second order autocorrelation function measured in HBT configuration, following the relationship\cite{Rezai2018}: $\abs{g^{(1)}(0)}^2=1-2g_{HBT}^{(2)}(0)=0.94 \pm 0.04 $ . It is important to notice that such visibility corresponds to the post-selected photons colliding at the beam splitter within a time-window much smaller than the photon wave packet duration. As such, it is effective only at the price of a reduced emitter brightness. In other words, suppressed coincidences at zero time delay are expected also in a HOM experiment with detuned sources or in presence of dephasing, with spectral fluctuations smaller than the inverse of the overall setup temporal resolution\cite{Lettow2010a}.
A non-perfect visibility in case of severe post-selection can be hence ascribed only to the detector time-jitter, or to optical misalignment of the setup. The HOM interference profile in CW operation still gives some insight about the indistinguishability of the considered photon stream. Indeed, while the minimum of the dip is related to purity and other technical aspects
, the characteristic time of the exponential profile depends on the coherence of the emission\cite{Rezai2018}. In details, we find from the fits that the time scale of the dip for parallel configuration is shorter ($\tau_{\parallel}=2.7 \pm 0.2$ $ns$) than that for orthogonal polarization and in the HBT case ($\tau_{\bot}=3.6(5) \pm 0.3(0)$ $ns$, $\tau_{HBT}=3.5 \pm 0.2$ $ns$). This mismatch is expected in presence of pure dephasing, and corresponds to an estimation of the coherence time of the emitter equal to $\tau_c=2\tau_{\parallel}=5.4 \pm 0.5$ $ns$. It is interesting to compare this value with the direct measurement of the ZPL linewidth obtained via excitation spectroscopy, that means recording the fluorescence signal in the PSB as a function of the excitation laser frequency, scanning across the ZPL (see Fig.\ref{fig1}b). Indeed, in the latter case a $FWHM$ of the lorentzian fit to data yields a value of $=55.1\pm 0.5$ $MHz$, compatible with the estimate from HOM interference yielding $FWHM= 1/ (\pi\tau_c)=59\pm 10$ $MHz$. We can conclude that the system shows negligible spectral diffusion under CW pumping, as the two estimate for the dephasing broadening are consistent within the error bars, although sensitive to frequency fluctuations on very different time scales ($40$ $ns$ in the HOM measurements and seconds-long integration time for the excitation spectroscopy).\\

\begin{figure*}
\includegraphics[width=\textwidth]{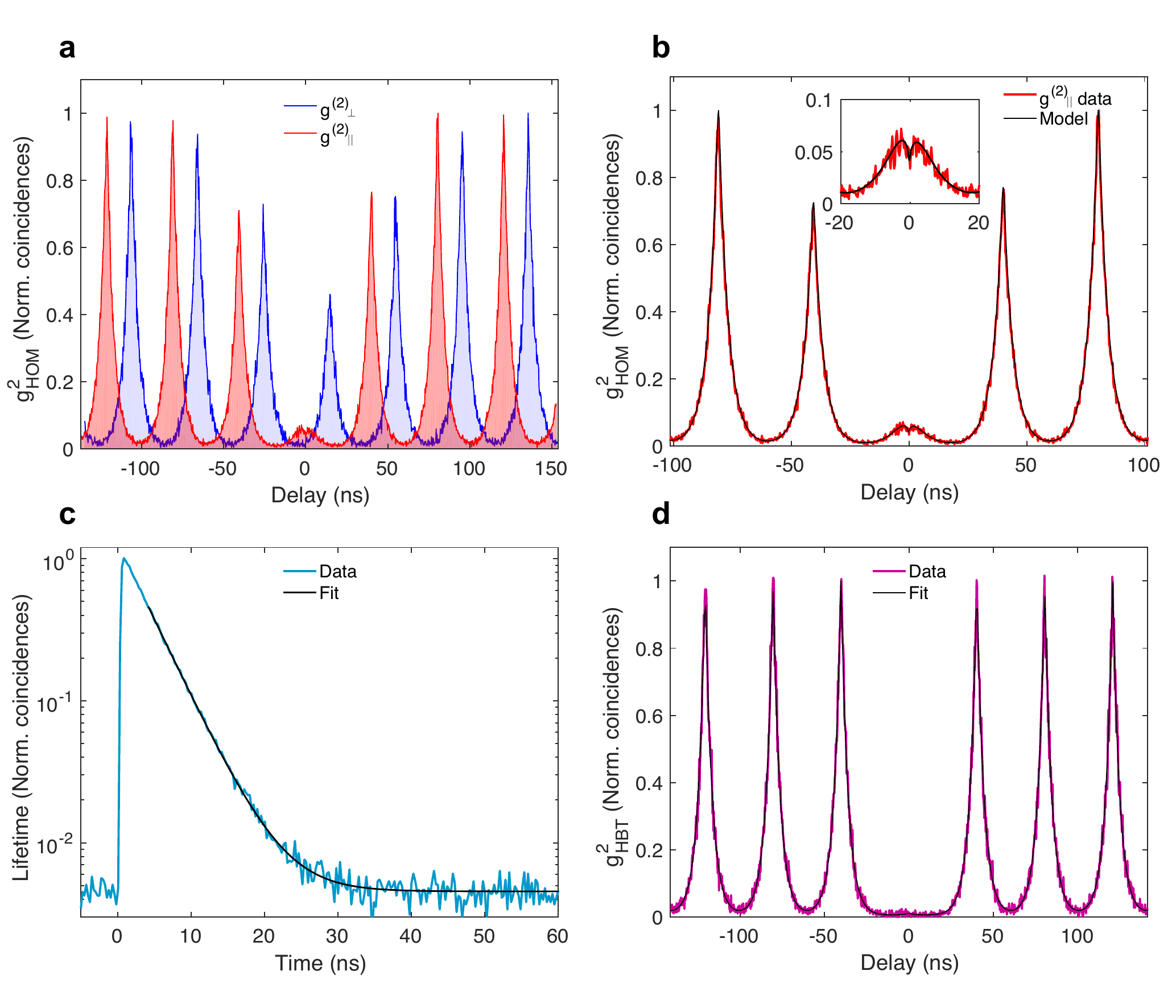}
\caption {\textbf{Characterization of the source under pulsed excitation. a:} Histogram of the relative arrival time for HOM setup in case of parallel (red) and orthogonal (blue) polarization of the input modes. The latter is shifted in time for clarity. \textbf{b:} Zoom around zero time delay for parallel polarization. \textbf{c:} Lifetime measurement and \textbf{d:} histogram approximating the $g^{(2)}(\tau)$ in the HBT configuration. Black solid lines in b, c, and d are fits to the data according to the equation in the main text.}
\label{fig2}
\end{figure*}
A more direct estimate of the photon wave packet overlap can be obtained addressing with laser pulses 
the molecule previously characterized in CW. First the fluorescence lifetime is measured, recording the histogram of the photon arrival times with respect to the laser excitation pulse.
A typical data set is reported in Fig.\ref{fig2}c and from the best fit with a single exponential decay, a lifetime $\tau_{1}=4.01\pm 0.01$ $ns$ is obtained, with uncertainty given by the standard deviation over $10$ measurements\footnote{In this case we can approximate the population relaxation time just with its radiative component, based on the characteristic molecule off-times in the triplet state, discussed e.g. in Ref.\cite{Nicolet2007a}}. In order to perform HOM interference
, the MZI delay and the temporal distance among neighbouring pulses have to coincide. To this aim, the repetition rate is adjusted according to the result for $\Delta t$ 
from the HOM fit in CW. The uncertainty associated to this measurement is estimated from the error in the fit parameter and amounts to about $ 200$ $ps$.
This directly impacts the uncertainty on the visibility, as it will be shown later. In Fig.\ref{fig2}a, the $g^{(2)}_{HOM}(\tau)$ under pulsed excitation are compared for parallel and orthogonal polarization in red and blue solid lines, respectively, with the second one artificially delayed for sake of clarity. All curves are normalized with respect to the maximum peak amplitude, calculated averaging over more than ten pulses, excluding the ones in zero and $\pm \Delta t$. The latter pulses are associated to coincidences between photons that are one or two periods apart, leading to areas $A_{k=-1}=N(1-R^2)$, $A_{k=+1}=N(1-T^2)$, whereas all other peaks can be due to 3 classes of events which sum up to a constant probability, $ A_{k\leq-2}=
A_{k\geq 2}=N
$, with $N$ being proportional to the integration time 
(suppl. Ref.\cite{Loredo2016}) 

One way of defining the TPI visibility is through the comparison of the peak area around zero delay for the case of indistinguishable versus distinguishable photons, calculating the ratio $V=\frac{A_{\bot}-A_{\parallel}}{A_{\bot}}$. In our case, integrating within a time window $\Delta T=26$ $ns$ (corresponding to about $96\%$ of the photon wave packet area), we estimate a visibility $V=78 \pm 4\%$. The results are further analysed with a model based on Ref.\cite{Kambs2018}, extrapolated for a train of infinite pulses (for details see SI), with the introduction of a phenomenological parameter $v$. This factor takes into account a non-perfect spatial alignment and polarization control, the residual multiphoton probability, as well as the effect of emission of distinguishable photons within the filter range but outside the ZPL. It amounts to the visibility that would be measured in the limit of negligible dephasing. 
As it can be appreciated in Fig.\ref{fig2}b, the model correctly describes the data, considering the 
unbalance between $\abs{r}^2$ and $\abs{t}^2$ of the second beam splitter, the temporal resolution of the electronic system ($230$ $ps$, obtained as standard deviation of the $g^{(2)}_{HBT}(\tau)$ for laser pulses, details in SI) and a delay given by the inverse of the laser nominal repetition rate ($24.79$ $MHz$). The only free parameters in the model are the excited state lifetime $\tau_1$, the pure dephasing $\Gamma^*$ and $v$. A least square fitting algorithm yields the following values: $\tau_1=4.04\pm 0.02$ $ns$, $\Gamma^*=55 \pm 10$ $MHz$ and $v=0.95 \pm 0.02$. The first two parameters are consistent with the independent measurement discussed before, in particular estimating the linewidth as $FWHM=\Gamma/\pi=1/(2\pi\tau_1)+\Gamma^*/\pi = 57\pm 4$ $MHz$.

Interestingly, the factor $v$ is the visibility that could be obtained just cooling the system further to $1.4$ $K$ \cite{Nicolet2007a}, in order to get truly lifetime-limited linewidths $(FWHM=1/(2\pi\tau_1)\simeq40$ $MHz)$. 
It is therefore important to understand what is limiting $v$ to a value smaller than 1. Although the polarization optics have been characterized with an extinction ratio of $1/1600$, the experimenter introduces an error in setting the relative orientation between the two arms. 
Another contribution could stem from the not-perfect purity of the single photon Fock state. This can be inferred from a measurement of the $g^{(2)}_{HBT}(\tau)$ in the HBT, reported e.g. in Fig.\ref{fig2}d, yielding $\Tilde{g}^{(2)}_{HBT}(0)=A_0/A_N= 0.008 \pm 0.008$. Based on Ref.\cite{ollivier2021hong}, we then estimate the mean wave packet overlap of the single photon component $M_S$, in case of negligible dephasing and assuming full distinguishability for the rest of the input state\cite{ollivier2021hong}, as $M_{S}=\frac{v+1}{4RT(1-\Tilde{g}^{(2)}(0))}-1\simeq 97\%$. Such value obviously depends on the filtering efficiency. In fact, the presence of a residual component of distinguishable photons $(1-\alpha)$, given by the portion of the emission in the PSB leaking through the notch filter, degrades the expected maximum visibility to\cite{Clear2020} $\alpha^2$. According to the fluorescence spectrum measured before and after the filter (reported in SI), roughly $98\%$ of the light overlaps with the ZPL. 
\begin{figure*}
\includegraphics[width=\textwidth]{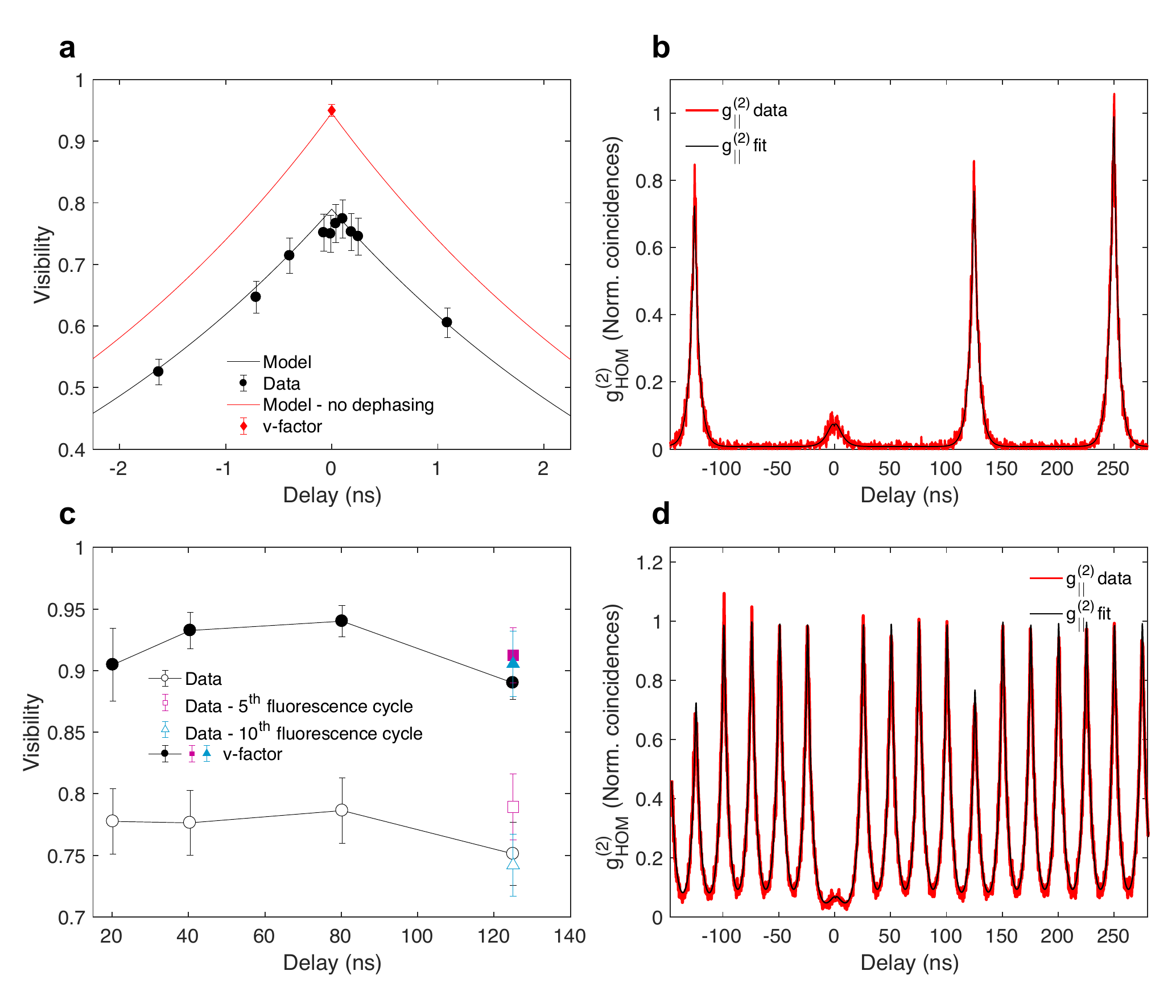}
\caption{\textbf{HOM visibility under pulsed operation. a:} visibility and $v$-factor as a function of the laser repetition rate. The delay is actually obtained varying the laser repetition rate, which determines the temporal overlap between the  photon wave packets. \textbf{b:} Coincidences histogram for $125$ $ns$-long ($25$ $m$) delay line and parallel polarization. \textbf{c:} optimal HOM visibility and $v$-factor versus delay line length; for longest delay ($125$ $ns$), also the visibility of the interference between photons separated by multiple fluorescence cycles is reported as colored dots. \textbf{d:} Coincidences histogram for $15$ $ns$-long delay line and parallel polarization, in case of $40$ $MHz$ repetition rate (interference between photons separated by five fluorescence cycles.)}
\label{fig3}
\end{figure*}
In the optimal condition of zero dephasing but with the same filtering window, one can hence expect a visibility of $96\%$, which is just what has been estimated above.

Scanning the laser repetition rate and hence effectively the delay between the two photon wave packets, different sets of data are recorded, and the corresponding values for the visibility are plotted in Fig.\ref{fig3}a, together with the derived theoretical curve. Using our model to estimate the error on the visibility, we observe that the main contribution to its uncertainty is given by the error in estimating the delay line ($\pm 0.2 ns$), that determines an uncertainty on V equal to about $4\%$.
TPI has been measured also as a function of the delay in the unbalanced MZI and the distance among interfering particles in number of photons. In Fig.\ref{fig3}b, the $g^{(2)}_{HOM}(\tau)$ using parallel polarizations for a delay of $125.0\pm0.2$ $ns$ (around $25$ $m$) is reported. 
It is worth noticing that the visibility starts to decrease only around such time delay (see Fig.\ref{fig3}c), which corresponds to about 30 times the photon wave packet extension. Remarkably, as reported in the visibility graph in Fig.\ref{fig3}c, we observe no relevant drop in TPI also between photons separated by up to $10$ fluorescence cycles. As an example, Fig.\ref{fig3}d shows the coincidences histogram for the case of interference between photons separated by 5 fluorescence cycles. Here a delay $\Delta t$ of $125$ $ns$ was employed with a laser repetition rate of $40$ $MHz$. Such a remarkable spectral stability is in agreement with the negligible spectral diffusion reported for CW operation both here and in literature\cite{pazzagli2018,Lettow2010a}.\\

In conclusion, in this paper we  demonstrate highly indistinguishable single photons obtained on demand from a single organic dye molecule under non-resonant excitation. The results are obtained for emitters in a sub-micrometric environment, without the help of any photonic resonance, and using only a 0.4 nm-wide spectral filter to select the emission. A HOM interference visibility of more than 75\% is reported, limited by the residual dephasing present at the operating temperature of 3 K. A visibility of 96\% is expected for the very same experiment in case of 1.5 kelvin operation. We also find that HOM visibility does not show relevant reduction for photons separated by up to 125 ns (equivalent to 30 times the wave packet duration) and by up to 10 fluorescence cycles. Such a spectral stability is of major interest for applications involving multiple photons, such as linear optical quantum computing, where temporal demultiplexing is a typical strategy to increase the number of available resources. The source presented in this paper shows a brightness at detector limited to around $2\%$, corresponding to a brightness at first lens (N.A.=0.67) of around 5\%. Hence, in the perspective of implementation for quantum applications, integration of the emitter with photonic devices becomes mandatory. A photonic resonance can modify both the radiation pattern and the spectral distribution of the emission, in order to bring the source brightness to the state-of-the-art level\cite{wang2019towards}. In this respect, the same type of system has been shown to be particularly suitable for the integration in hybrid photonic structures\cite{Colautti2020,Lombardi2017a}.

\section*{Acknowledgements}
The authors would like to thanks Prof. Christoph Becher and Dr. Benjamin Kambs for fruitful discussion. This project has received funding from the EraNET Cofund Initiatives QuantERA
within the European Union's Horizon 2020 research and innovation program grant agreement No. 731473 (project ORQUID)  and from the 
 EMPIR programme (project 17FUN06, SIQUST) co-financed by the Participating States and from the European Union’s Horizon 2020 research and innovation programme.

\bibliographystyle{qute}
\bibliography{publicationsjabRef2}

\end{document}